\documentclass[a4paper,11pt]{article}
\usepackage{pos}

\usepackage{enumitem}
\usepackage{caption}
\usepackage{subcaption}
\usepackage[
 print-unity-mantissa=false,
 exponent-product=\ensuremath{\cdot},
 uncertainty-mode = separate,
]{siunitx}

\hypersetup{
    pdftitle={SAG-SCI: the Real-time, High-level Analysis Software for Array Control and Data Acquisition of the Cherenkov Telescope Array Observatory},
    pdfauthor={Gabriele Panebianco},
    pdfsubject={Real-time High-level Analysis Software for the CTAO},
    pdfkeywords={real-time, gamma-rays, software},
    pdfcreator={LaTeX with hyperref},
    pdfproducer={pdflatex}
}
\title{SAG-SCI: the Real-time, High-level Analysis Software for Array Control and Data Acquisition of the Cherenkov Telescope Array Observatory}
\ShortTitle{SAG-SCI: Real-time, High-level Analysis Software for the CTAO}

\author*[a]{Gabriele Panebianco}
\author[a]{Nicolò Parmiggiani}
\author[a]{Andrea Bulgarelli}
\author[a]{Ambra Di Piano}
\author[a]{Luca Castaldini}
\author[a]{Valentina Fioretti}
\author[a]{Giovanni De Cesare}
\author[b]{Sami Caroff}
\author[b]{Pierre Aubert}
\author[b]{Gilles Maurin}
\author[b]{Vincent Pollet}
\author[b]{Thomas Vuillaume}
\author[c]{Igor Oya}
\author[d,a]{Cristian Vignali}
\onbehalf{for the CTAO-ACADA Collaboration}

\affiliation[a]{INAF - Osservatorio di Astrofisica e Scienza dello spazio di Bologna,\\Via Piero Gobetti 93/3, 40129 Bologna, Italy}
\affiliation[b]{Univ. Savoie Mont Blanc, CNRS, Laboratoire d'Annecy de Physique des Particules - IN2P3\\74000 Annecy, France}
\affiliation[c]{Cherenkov Telescope Array Observatory\\ Saupfercheckweg 1, D-69117, Heidelberg, Germany}
\emailAdd{gabriele.panebianco@inaf.it}
\affiliation[d]{Dipartimento di Fisica e Astronomia (DIFA) Augusto Righi, Università di Bologna,\\Via Piero Gobetti 93/2, I-40129 Bologna, Italy}

\abstract{
The Cherenkov Telescope Array Observatory (CTAO) is going to be the leading observatory for very-high-energy gamma-rays over the next decades.
Its unique sensitivity, wide field of view, and rapid slewing capability make the CTAO especially suited to study transient astrophysical phenomena.
The CTAO will analyse its data in real-time, responding to external science alerts on transient events and issuing its own.
The Science Alert Generation (SAG) automated pipeline, a component of the Array Control and Data Acquisition (ACADA) software, is designed to detect and issue candidate science alerts.

In this work, we present the current development status of SAG-SCI, the SAG component responsible for the real-time, high-level analysis of CTAO data.
The SAG-SCI pipelines receive gamma-ray data from multiple reconstruction lines, merge them, store them in a database, and trigger several parallel scientific analyses on the latest data.
These analyses include estimating target significance and flux, producing sky maps and light curves, and conducting blind searches for sources within the field of view.
We execute SAG-SCI on a set of simulated gamma-ray data, detecting the simulated sources and accurately reconstructing their flux and position.
We also estimate the systematic errors introduced by the analysis and discuss the results in relation to the generation of candidate science alerts.
}

\ConferenceLogo{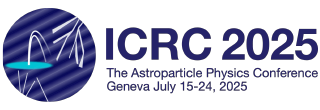}

\FullConference{39th International Cosmic Ray Conference (ICRC2025)\\
 15–24 July 2025\\
Geneva, Switzerland\\}

\begin{document}
\maketitle
\section{Introduction}
The Cherenkov Telescope Array Observatory (CTAO) will be the most sensitive instrument for very-high-energy (VHE) gamma-ray astrophysics in the coming decade and beyond \cite{McMuldroch_CTAO_2024}.
The analysis and interpretation of gamma-ray data are crucial for understanding transient phenomena.
In fact, only the most extreme environments and processes in the Universe, such as cosmic-ray acceleration sites and compact objects like neutron stars and black holes, exhibit the dynamic processes and energy budget required to produce high-energy photons, neutrinos and gravitational waves.
They are most suited to emit them during violent transient events.
The CTAO is therefore expected to play a pivotal role in time-domain, multi-wavelength and multi-messenger astrophysics, owing to its broad energy range, wide field of view (FoV), fast telescope slewing capabilities and unprecedented sensitivity for short timescales.
For these reasons, the study of transients represents one of the CTAO's key science projects.
To maximise its scientific return, the CTAO will feature a low-latency, effectively real-time analysis pipeline as part of its Array Control and Data Acquisition (ACADA) system \cite{oya_acadaspie_2024}, called the Science Alert Generation (\verb|SAG|) pipeline \cite{bulgarelli_SAGproceedingsSPIE_2024}.

The main goals of the \verb|SAG| pipeline are to reconstruct Cherenkov data acquired by the CTAO, perform data quality checks to identify potential issues, analyse the data and enable early identification of transient phenomena.
Each task is handled by a dedicated \verb|SAG| subsystem: \verb|SAG-SUP| supervises the pipeline execution \cite{Castaldini_SUP_ICRC2025}, \verb|SAG-RECO| performs the Cherenkov reconstruction \cite{caroff_reco_2023} and \verb|SAG-DQ| conducts data quality checks \cite{baroncelli_dq_2022}.
Finally, the analysis of reconstructed data is carried out by the \textquotedblleft High-Level Analysis Pipeline\textquotedblright, or \verb|SAG-SCI|.
\verb|SAG-SCI| has two primary objectives: (1) to perform continuous scientific monitoring of the observed sources, and (2) to generate candidate science alerts when transient events are detected.

A science alert is a scientific communication indicating the occurrence of a transient phenomenon, shared as rapidly as possible among observatories and research facilities to facilitate multi-wavelength follow-up observations and enhance the scientific understanding of the event.
Common examples include Astronomer's Telegrams (ATels) and notices or circulars of the General Coordinate Network (GCN).
Through the ACADA system and the \verb|SAG| pipeline, the CTAO will not only react to alerts from external facilities, swiftly repointing its telescopes to observe the events, but will also issue internal science alerts to the broader scientific community.
Specifically, \verb|SAG| will issue internal candidate science alerts, which will be confirmed or rejected by the ACADA Transient Handler subsystem.

In this work, we present the current status of \verb|SAG-SCI|: section~\ref{sec:software} describes the software requirements, design and typical workflow, while section~\ref{sec:analyses_alerts} describes the scientific analyses implemented and the strategies to generate candidate science alerts.
Section~\ref{sec:validation} describes the validation of \verb|SAG-SCI| in several simulated observational scenarios.
We discuss our results in section~\ref{sec:conclusion}.
A more in-length discussion of \verb|SAG-SCI| is reported in \cite{panebianco_phdthesis}.

\section{Software Requirements and Design}
\label{sec:software}

\verb|SAG-SCI| is designed to analyse every scientific observation run of the CTAO and must satisfy strict requirements to ensure product and result quality.
A core requirement is analysis flexibility and completeness: \verb|SAG-SCI| will provide the statistical significance and the estimated flux of the observed sources, implementing different types of analyses, including light curve estimation, computation of sky maps, full-FoV blind searches to detect known and unknown sources.
Analysis timescales used to detect transients can vary from $\SI{10}{\second}$ to $\SI{180}{\minute}$.
In terms of computational performance, \verb|SAG-SCI| must be capable to perform different analyses in parallel and prioritize urgent tasks, giving precedence to high-priority analyses such as transient searches, especially for transients observed during Target of Opportunity (ToO) observations as follow-up of science alerts.
Finally, \verb|SAG-SCI| must issue candidate science alerts within $\SI{5}{\second}$ of data availability (latency requirement), and suppress or flag science alerts when data quality warnings or alarms are present.

These software requirements drive the design and architecture of \verb|SAG-SCI|, which is implemented as \verb|python| package with a \verb|MySQL| interface, and based on the \verb|RTApipe| framework \cite{parmiggiani_rtapipe_2021}.
\verb|SAG-SCI| follows a modular, object-oriented approach to implement its main components:
\begin{itemize}[noitemsep, leftmargin=.15in]
    \item \textbf{Data Source}.
    A \verb|MySQL| database that stores input data acquired by the telescope sub-arrays, including physical properties of the $\gamma$-ray reconstructed events and metadata (e.g., file paths).
    \item \textbf{Data Model}.
    A \verb|MySQL| database that defines available observations, instruments, analysis types and targets.
    It includes both static elements (e.g., instruments and analysis types), and dynamic elements updated during operations (e.g., targets analysed and analysis jobs run).
    \item \textbf{Science Logic}.
    A set of \verb|MySQL| triggers that encode the rules for executing analyses as defined in the Data Model, in accordance with software requirements and the use cases.
    \item \textbf{Pipeline Manager}.
    A set of $3$ \verb|python| daemons responsible for managing pipeline operations: inserting input $\gamma$-ray event files in the database (\verb|DL3Merger|), executing analyses when in a \textquotedblleft runnable\textquotedblright\ state (\verb|SubmitJob|), and monitoring their status and completion (\verb|UpdateJobStatus|).
    \item \textbf{Task Manager}.
    An external component responsible for allocating resources, scheduling and executing analysis jobs when prompted by the Pipeline Manager.
    The current version of \verb|SAG-SCI| uses the \verb|slurm| Workload Manager.
    \item \textbf{Science Tool Wrappers}.
    A set of classes and scripts performing analyses with scientific libraries.
    \verb|SAG-SCI| currently has a wrapper for \verb|gammapy| \cite{gammapy_paper_2023} and a custom \verb|python| wrapper, \verb|RTAPH| \cite{dipiano_RTAPH_2022}.
    \item \textbf{Results Database}.
    A \verb|MySQL| database that stores analysis jobs results and candidate alerts.
\end{itemize}

\begin{figure}[t!]
  \centering
 \makebox[\textwidth][c]{\includegraphics[width=0.99\textwidth]{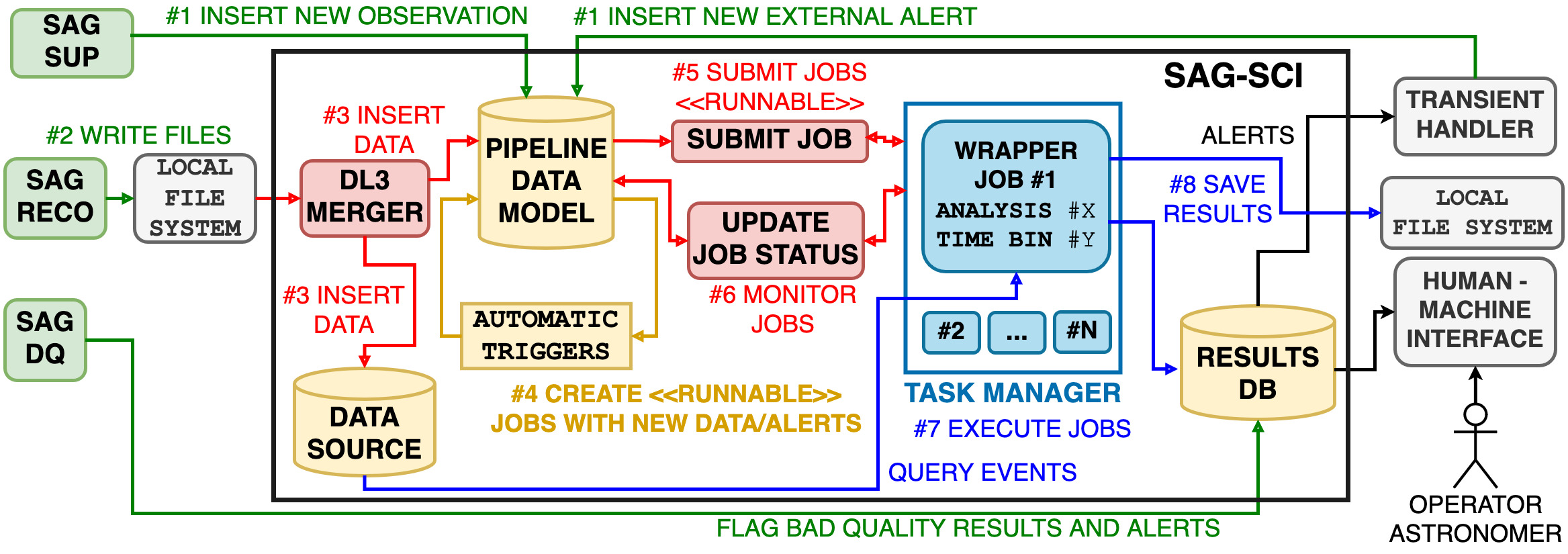}}
 \caption{
 Overview of SAG-SCI components and their interactions during the analysis workflow.
 SAG-SCI Pipeline Manager daemons are drawn in red, databases in yellow, Science Tools wrappers in blue.
 Other SAG subsystems are drawn in green, components external to SAG in grey.
 Numbers highlight the workflow described in section~\ref{sec:software}.
 }
 \label{fig:SCI_schema}
\end{figure}

\noindent In Figure~\ref{fig:SCI_schema} we show the steps performed by \verb|SAG-SCI| during its analysis workflow:
\begin{enumerate}[noitemsep, leftmargin=.15in]
    \item \verb|SAG-SUP| starts the pipeline and inserts information on a new run and its scientific target in the Data Model database.
    Alternatively, if a science alert is received, \verb|SAG-SCI| waits for the corresponding data to become available.
    \item \verb|SAG-RECO| generates $\gamma$-ray photon list files for the run and stores them in a dedicated directory.
    \item The \verb|DL3Merger| is notified of the new data, reads it and inserts it in the Data Source database.
    \item The insertion of new data triggers the creation of analysis jobs in the Data Model database according to the Science Logic rules.
    Each job analyses a specific time window within the observation run.
    \item The \verb|SubmitJob| periodically queries the Data Model database to check for runnable jobs.
    When a new job is ready, it is submitted to the Task Manger scheduler (\verb|slurm|).
    \item The \verb|UpdateJobStatus| registers the job submission, and may pause or cancel jobs if higher-priority tasks need to be executed.
    \item The Task Manager executes the job on the available computing resources using the configured Science Tool Wrapper.
    \item The Wrapper saves the results to both the file system and the Results database.
\end{enumerate}

\section{Scientific Analyses and Generation of Science Alerts}
\label{sec:analyses_alerts}
The primary outputs of the scientific analyses are the statistical significance of a target source and its average flux within a given time window.
\verb|SAG-SCI| schedules analysis jobs as soon as the data becomes available using a \textquotedblleft sliding-window\textquotedblright\ approach to construct a light curve of the source.
This method enables the detection of short-timescale variability and transient events.
In addition, previous results can be \textquotedblleft stacked\textquotedblright, i.e. combined with the most recent ones, to improve photon statistics and effective area, enhancing the detection of faint sources.
Stacked results yield the time-averaged flux and the cumulative significance of the source, and are especially important for scientific monitoring.
Parallel analyses can also be performed to generate sky counts and significance maps, compute the radial distribution of the photons around the source and a control region (the so-called $\theta^2$ plots), and carry out blind searches for unknown sources, poorly localised or serendipitous transient sources in the Field of View (FoV).
Blind searches are particularly important during surveys or when following up on poorly localised transients, such as gravitational-waves.

Candidate science alerts are issued when the estimated significance and flux exceed predefined thresholds.
Alerts are expected to be generated in the following main scenarios:
\begin{itemize}[noitemsep, leftmargin=.15in]
    \item \textbf{Detection of a Flaring State of a known source.}
    When a detected source matches an entry in a reference VHE catalogue, its measured flux is compared with the catalogue value.
    If the flux is significantly higher (or lower), the source is in a flaring (or lower) state, and a candidate science alert is issued.
    \item \textbf{Follow-up of external science alerts.}
    If the observed target is not present in a catalogue, as is often the case for one-time transients, e.g., gamma-ray bursts (GRBs), and ToO observations, an alert is generated upon detection.
    Catalogue cross-matching may help identify potential counterparts.
    For extra-galactic transients, flux estimation strongly depends on the redshift, due to absorption by the extra-galactic background light (EBL).
    Since redshift estimation typically requires multi-wavelength observations and is often unavailable in real-time, reliable flux estimates may not always be possible.
    \item \textbf{Transient detection using the Blind search algorithm.}
    The Blind search algorithm identifies hotspots of excess counts in the FoV.
    If a hotspot exceed the statistical significance threshold and is either unassociated with any known source or matches a known source in a flaring state, a candidate science alert is generated.
    In this case, the pipeline also register the source as a new target and initiates a dedicated analysis centred on its position.
\end{itemize}


\section{Validation with Simulated Observational Scenarios}
\label{sec:validation}
The scientific and technical performance of \verb|SAG-SCI| is evaluated through several automated test suites.
These include: unit tests to verify the correct execution of individual classes, functions and modules; verification tests to assess the end-to-end workflow of \verb|SAG-SCI| as a stand-alone software; integration tests to evaluate its behaviour as a sub-system within \verb|SAG| and ACADA.
In this work, we designed $3$ observational scenarios using CTAO simulated data to validate \verb|SAG-SCI|'s performance in relevant scientific use cases.
All simulations were performed with \verb|gammapy v1.2|, assuming an array of $4$ Large-Sized Telescopes at the CTAO-North site.
Observations were simulated in wobble-mode at a zenith angle of $\SI{20}{\deg}$, using the CTAO Prod$5$ Instrument Response Functions \cite{CTAO_ZenodoIRFs_2021}.
The analyses were performed with the \verb|gammapy| wrapper implemented in \verb|SAG-SCI|, in the $\qtyrange[range-phrase=-]{0.1}{10}{\tera\electronvolt}$ energy range.

\paragraph{Monitoring of a Steady Source.}
To verify the correct behaviour of the Science Tool Wrappers and emulate steady-source monitoring scenarios (e.g. for calibration or archival purposes), we simulated a point-like steady source with a power-law spectrum (spectral index $\Gamma=-2$) and a photon flux of $\phi=\SI{9.9E-11}{\second^{-1}\centi\meter^{-2}}$ in the simulated energy range.
The analysis used the $1$D \verb|gammapy| analysis approach, estimating the background via aperture photometry and evaluating the source flux through maximum likelihood fitting.
\verb|SAG-SCI| successfully detected the source, and estimated flux values consistent, within statistical uncertainties, with the simulated value, both in the light curve and the stacked average flux.
Typical statistical errors on the flux, driven by the fit, are $\approx 35\%$, mainly due to the limited photon statistics at short timescales.
Currently, \verb|SAG-SCI| estimates flux assuming a predefined spectral model (a power-law) fitting only the normalisation factor.
As a result, the estimation of the flux is sensitive to the assumed spectral index.
An estimation of the index in real-time is not always possible due to poor photon statistics.
To quantify this systematic error, we repeated the analysis with a spectral index offset by $20\%$, which introduced a $\approx 30\%$ systematic error in the flux estimation.
While other sources of systematic uncertainty (e.g., from Cherenkov reconstruction) exist, this is the dominant one introduced by \verb|SAG-SCI| itself.

\begin{figure}[t!]
  \centering
 \makebox[\textwidth][c]{\includegraphics[width=0.75\textwidth]{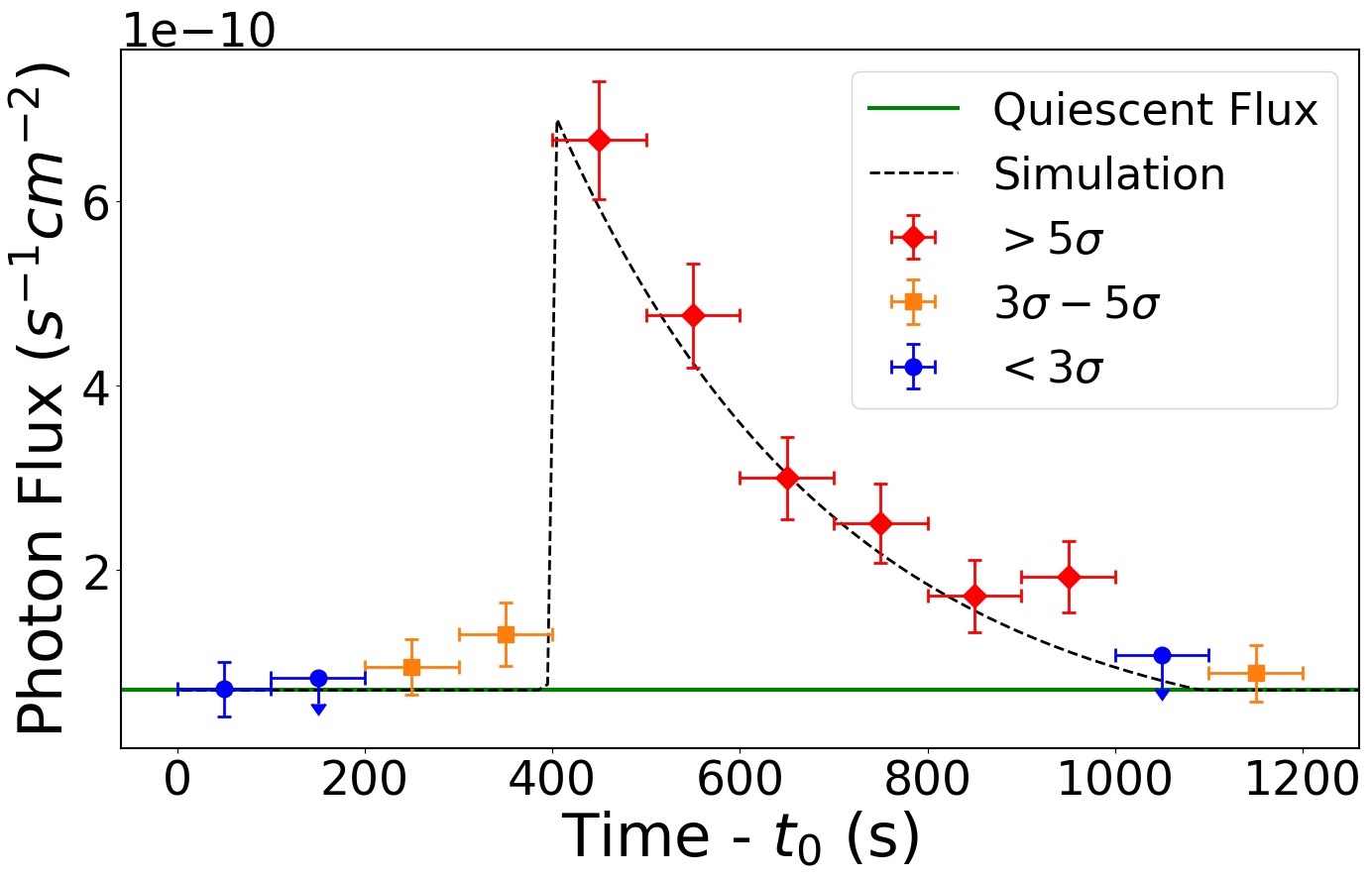}}
 \caption{
 Light curve of a simulated flaring source computed by SAG-SCI.
 Flux points are colour-coded for their statistical significance.
 The measured flux is in agreement with the simulated value (black dashed line).
 The flare is detected in the fifth bin, where the flux is not compatible with the quiescent reference value (green line), triggering a candidate science alert.
 }
 \label{fig:LC_flare}
\end{figure}

\paragraph{Detection of a Gamma-ray Flare.}
To simulate a $\gamma$-ray flare, we modelled a point-like source with a power-law spectrum ($[0.1-10]\,\si{\tera\electronvolt}$ flux $\phi=\SI{6.9E-11}{\second^{-1}\centi\meter^{-2}}$, index $\Gamma=-2$), modulated by a dimensionless function $f(t)$.
The modulation remains constant before the flare onset time $t_0$ $\left(f(t<t_0)=1\right)$, jumps to $f(t_0)=10$ at the onset, and then decays exponentially with a timescale $\tau=\SI{300}{\second}$.
The flare onset was simulated $\SI{400}{\second}$ after the start of the observation.
We emulated the progressive arrival of gamma-ray data and computed a $\SI{100}{\second}$ binned light curve using \verb|SAG-SCI|, shown in Figure~\ref{fig:LC_flare}.
The flare was successfully detected in the fifth analysis bin, with significance $>5\sigma$, and flux $\sim 10$ times greater than the quiescence reference value.
Candidate science alerts will be issued by \verb|SAG-SCI| when significant deviations from reference fluxes are observed.

\begin{figure}[htp]
 \centering
 \makebox[\textwidth][c]{\includegraphics[width=0.8\textwidth]{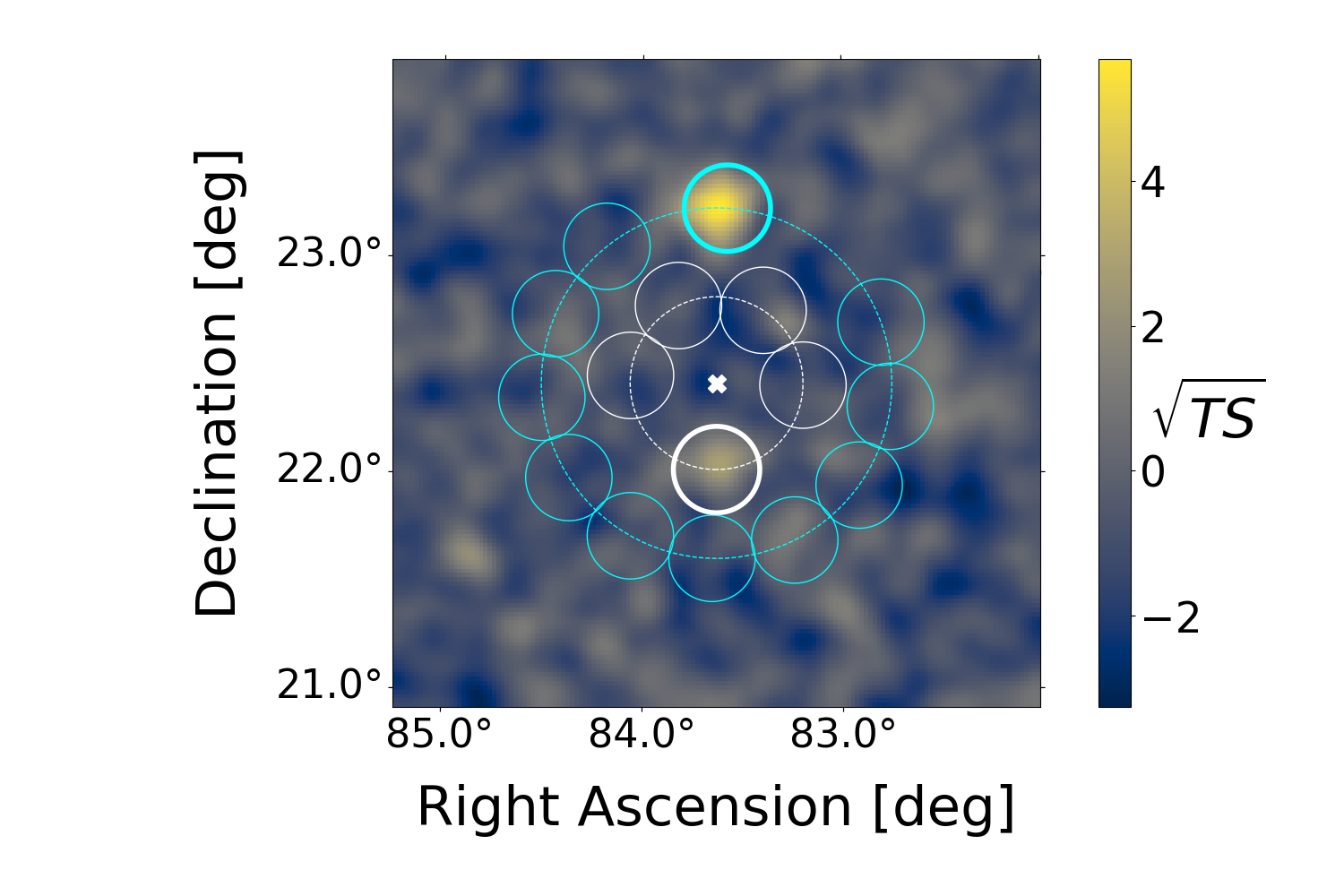}}
 \caption{
 Significance sky map (square root of the Test Statistic $TS$) of the third simulation.
 The white cross marks the sub-array pointing direction, the thick cyan circle the on-source region centred on the serendipitous transient detected by the Blind search.
 Its statistical significance ($6.9\sigma$) is computed with the reflected background regions method using symmetric off regions (thin cyan circles).
 We also show the position of the simulated steady source (thick white circle) and its off regions (thin white circles).
 Dashed circles mark the $\SI{0.4}{\deg}$ and $\SI{0.8}{\deg}$ offset.
 }
 \label{fig:BlindSearch}
\end{figure}

\paragraph{Detection of a Serendipitous Transient with the Blind Search Algorithm.}
To test the Blind search algorithm, we simulated a FoV with two sources: a steady, power-law, point-like source located at a $\SI{0.4}{\deg}$ offset, representing the primary science target ($[0.1-10]\,\si{\tera\electronvolt}$ flux $\phi=\SI{5.0E-11}{\second^{-1}\centi\meter^{-2}}$, index $\Gamma=-2$), and a transient source at $\SI{0.8}{\deg}$ offset, mimicking a serendipitous event.
The transient's model is a point-like source with power-law spectrum ($[0.1-10]\,\si{\tera\electronvolt}$ flux $\phi=\SI{5.0E-11}{\second^{-1}\centi\meter^{-2}}$, index $\Gamma=-2$) modulated by an exponential decay function.
We used the same function of the gamma-ray flare, with the only difference being the flux being zero before the transient onset.
\verb|SAG-SCI| performed stacked analyses every $\SI{100}{\second}$ of received data, producing sky maps with increasing exposure and running the Blind search algorithm to identify excesses.
Figure~\ref{fig:BlindSearch} shows the sky map at $\SI{500}{\second}$, highlighting both the nominal position of the steady source targeted by the analysis and the newly detected transient.
The Blind search and the regular light curve analysis of the steady source were executed in parallel.
The detection of a new source will automatically trigger dedicated analyses, to estimate the light curve and the average flux of the new source, using both past and incoming data.

\section{Discussion and Conclusion}
\label{sec:conclusion}
We presented \verb|SAG-SCI|, the software developed for real-time gamma-ray data analysis and candidate science alert generation for the \verb|SAG| pipeline, which is a component of ACADA system of the CTAO.
It operates through $3$ \verb|python| daemons (the Pipeline Manager), handling data collection and job submission.
All input data, pipeline metadata, and results are stored in dedicated \verb|MySQL| databases.
Analysis jobs are managed by the \verb|slurm| Workload manager and performed with a \verb|gammapy| wrapper.
\verb|SAG-SCI| employs a sliding-window approach to produce real-time light curves as data becomes available, with the option to stack results for time-averaged flux and significance estimates.
Parallel jobs generate sky maps, $\theta^2$ plots and perform blind search of sources within the FoV.
\verb|SAG-SCI|'s modular, object-oriented architecture supports flexibility, testing and deployment.

In addition to the ongoing automated test suites, we validated \verb|SAG-SCI| with $3$ simulated observational scenarios.
The first scenario, the monitoring of a steady source, allowed us to verify the correct behaviour of the pipeline.
The estimated flux is consistent with the simulated value within statistical errors of $\approx 35\%$.
We also estimated that a systematic error of $\approx 30\%$ can be added to consider a possible $20\%$ error on the assumed spectral index.
The second scenario reproduced the detection of a gamma-ray flare.
\verb|SAG-SCI| successfully detected the flare, recovering both the timing and the flux, thus demonstrating its capability to trigger alerts when flux deviates from reference values.
In case where the reference flux is not available, the generation of candidate alerts can be prompted just if the source is detected.
This may happen for some high-priority transients, such as GRBs.
Finally, the third simulated scenario showed the detection of a serendipitous, off-target transient in the FoV using the Blind search algorithm.
This highlight the pipeline's ability to operate parallel analyses and dynamically detect and track new sources alongside the primary target.

A first version of \verb|SAG-SCI| was released with \verb|ACADA v1.0|, and deployed during the October 2023 test campaign on the CTAO Large-Sized Telescope Prototype (LST$-$1), where it successfully detected the Crab Nebula \cite{acada_testlst1_2023}.
Future versions will introduce key enhancements.
One key advancement will be increasing complexity of the input source models, leveraging the flexibility of the \verb|gammapy| library to support more realistic and detailed astrophysical scenarios.
Another important development will be the implementation of a real-time background model estimation capability, designed to provide robust performance for sky map estimation, even in cases where the theoretical or simulated models are not applicable.
Finally, the system will support full cross-checking against gamma-ray and multi-wavelength source catalogues, further improving the reliability and scientific value of candidate alerts by enabling contextual enrichment and consistency verification.

\verb|SAG| represents a significant advancement for CTAO's real-time science capabilities, supporting both routine monitoring and time-critical response to transient events.
The \verb|SAG| pipeline will transform high volumes of gamma-ray data into scientifically relevant outputs within seconds.
It will allow the CTAO to become a powerful observatory for time-domain astrophysics, and enable multi-wavelength and multi-messenger approaches that will lead to a deeper understanding of the broad-band non-thermal properties of target sources.

\section*{Acknowledgements}
\noindent We gratefully acknowledge financial support from the agencies and organizations listed here:

\noindent \href{https://www.ctao.org/for-scientists/library/acknowledgments/}{https://www.ctao.org/for-scientists/library/acknowledgments/}


\end{document}